\documentclass{llncs}
\usepackage{makeidx}  % allows for indexgeneration
\usepackage{graphicx} % for pdf, bitmapped graphics files
 \usepackage[caption=false,font=footnotesize]{subfig}
\usepackage[cmex10]{amsmath} % assumes amsmath package installed
\usepackage{amssymb}  % assumes amsmath package installed
\usepackage{color}
\usepackage{bbm}
\usepackage{algorithm,algorithmicx,algpseudocode}
\usepackage{url}

\usepackage{float}

\newcommand{\firenze}{City\#3}
\newcommand{\copenhagen}{City\#4}

\newcommand{\muenchen}{City\#6}

\newcommand{\stockholm}{City\#8}
\newcommand{\torino}{City\#9}
\newcommand{\wien}{City\#10}

\begin{document}

%
%\mainmatter              % start of the contributions
%
\title{Car sharing through the data analysis lens}
\titlerunning{Car sharing through the data analysis lens}  % abbreviated title (for running head)
\author{Chiara Boldrini \and Raffaele Bruno \and Haitam Laarabi}
\authorrunning{Ivar Ekeland et al.} % abbreviated author list (for running head)
\institute{IIT-CNR, Pisa, Italy\\
\email{first.last@iit.cnr.it}}

\maketitle              % typeset the title of the contribution

\begin{abstract}
Car sharing is one the pillars of a smart transportation infrastructure, as it is expected to reduce traffic congestion, parking demands and pollution in our cities. From the point of view of demand modelling, car sharing is a weak signal in the city landscape:  only a small percentage of the population uses it, and thus it is difficult to study reliably with traditional techniques such as households travel diaries. In this work, we depart from these traditional approaches and we rely on web-based, digital records about vehicle availability in 10 European cities for one of the major active car sharing operators. We discuss how vehicles are used, what are the main characteristics of car sharing trips, whether events happening in certain areas are predictable or not, and how the spatio-temporal information about vehicle availability can be used to infer how different zones in a city are used by customers. We conclude the paper by presenting a direct application of the analysis of the dataset, aimed at identifying where to locate maintenance facilities within the car sharing operational area. 
\end{abstract}
%ba
\section{Introduction}
\label{sec:intro}
\noindent
%[The role of car sharing in smart transportation]
%
Automobile transportation has been one of the main drivers of the population growth and increasing wealth that have characterised the last two centuries~\cite{mitchell2010reinventing}. Thanks to cars, people have had greater access to jobs, goods, services. These benefits have not come for free. The price paid for our increased mobility has been huge in terms of environmental pollution, city congestion and resulting health issues. We are now at a turning point for personal mobility systems: policy makers and citizens share the common idea that it is time to rethink the way we move. There are three main driving forces behind this personal mobility revolution: smart transportation, sharing economy, and green vehicles, all being tightly intertwined. The move from ownership mindset to usage mindset will make it possible to have significantly fewer vehicles in our cities. The implications are that we can save space (public parking space and private garage space) and use it for something with increased added vehicle than to host idle cars for hours (a private car is used only $5\%$ of its available time, corresponding to 72 minutes per 24 hrs). This usage mindset will allow people to use the car size most appropriate to their daily needs. Since the average vehicle has only $1.3$ occupants, people can refrain from buying a car able to address the extreme case of personal mobility (e.g., moving a whole family for a vacation), and instead use two-seaters that are more suitable for everyday commuting. On occasion, depending on their needs, they can rent larger vehicles, thus implementing the Mobility-as-a-Service concept. The, the virtuous cycle is completed with the switch to electric vehicles, which allow for a drastic reduction of the personal mobility carbon footprint.

% pippone solito sul car sharing
Car sharing is one of the pillars of the smart transportation concept, with its potential to reduce parking needs, promoting both public transport and non-motorised transportation modes (biking and walking), thus lowering households' transportation costs and traffic congestion in our cities~\cite{shaheen2015mobility}. The general idea of car sharing is that the members of a car sharing system can pick up a shared vehicle of the car sharing fleet when they need it. Different operators may implement different pickup/drop-off policies. In station-based systems, members can only pick and drop vehicles at designated locations called stations, as in the Autolib system in Paris. If the service is two-way (e.g., Zipcar, Modo), people are asked to bring back the vehicle to the station where they initially picked it up. Otherwise, the service is called one-way. One-way services are definitely the most popular among customers thanks to the flexibility they provide. Examples of one-way car sharing are Autolib, Ha:Mo ride, CITIZ. One-way services can drop altogether the concept of station: this is the case of so-called free floating car sharing -- such as Car2go, DriveNow, Enjoy -- whose customers can pick up and drop off vehicles anywhere with a defined operation area. 

%SoA on CS & data
Car sharing is a \emph{weak signal}  in the city landscape: the fraction of people relying on car sharing for their daily trips is rapidly increasing but it is still in the order of single digit percentage points in the best cases~\cite{kortum2014driving}. For this reason, so far car sharing has been mostly studied through surveys and direct interviews with its members~\cite{schwieger2015global,shaheen2015mobility}. In addition, car sharing is typically not accounted for in households travel diaries periodically collected by city administrations. Even if it were, the limitations of travel surveys are widely acknowledged, and range from their inability to capture changes in the routine travel behaviour to their underestimation (because of underreporting from people) of short, non-commute trips. Moreover, running a survey is very expensive if one wants to capture a statistically meaningful sample. 

%[This is even worse if we concentrate on the system perspective rather than on the customers' perspective. EXAMPLES]

% the new approach we take in this paper
Cities have been considered kaleidoscopes of information since a long time~\cite{meier1962communications} but the extent to which this is true has reached new heights now that a myriad of electronic devices have weaved into its fabric. From the car sharing perspective, this means that we can now know exactly when and where cars are available, and we can observe shared vehicle flows \emph{as they happen} in the city. This knowledge opens up a new avenue of research that goes in the direction of the new science of cities and urban computing: using data and electronic devices to extract knowledge and to improve urban solutions. Along these lines, the goal of this paper is to stimulate a discussion on how to apply urban computing ideas to the car sharing domain. To this aim, we exploit the availability of data about free floating car sharing in 10 European cities and we carry out an analysis with the following objective in mind: to understand what mining this kind of data can bring to cities and to car sharing operators.

\section{Related work}
\label{sec:relwork}
\subsection{Knowledge mining for car sharing}

Until recently, knowledge about car sharing systems has been mostly acquired through surveys~\cite{shaheen2015mobility,schwieger2015global}, in which car sharing operators and members are interviewed. The understandings and advancements brought about by these works are invaluable, but the collection of survey data is expensive, time consuming, and does not scale. For these reasons, in this work we depart from this approach and we exploit public, web-based, digital records, whose geotagged and time-stamped variety of data can be analysed employing data mining techniques.

Data-driven analysis of car sharing systems has been carried out in \cite{Willing2017,boldrini2016characterising,Kortum2016,Schmoller2015,Muller2015}. Both Schm{\"o}ller et al.~\cite{Schmoller2015} and this paper investigate car sharing usage and factors that may influence the demand for car sharing. Schm{\"o}ller et al.~\cite{Schmoller2015} carry out their analysis using a dataset provided by the car sharing operator, which contains more information than what is generally available to the research community at large. The analysis presented in this paper is instead based on data that are publicly available and it proves that publicly available data can already offer many insights in the car sharing operations. In addition, while Schm{\"o}ller et al.~\cite{Schmoller2015} focus on two cities (Munich and Berlin), here we consider several cities and we aim at finding invariants and dissimilarities between them. In this sense, our work is close to~\cite{Kortum2016}, which considers free-floating car sharing in multiple cities.  However, Kortum et al.~\cite{Kortum2016} focus on the growth rate of free floating car sharing rather than on the characterisation from the supply side point of view. Willing et al.~\cite{Willing2017}, using data from free floating car sharing in Amsterdam and Berlin tackles the problem of understanding if Points of Interest (PoI) in each city can be used as predictors of car sharing demand. M{\"u}ller and Bogenberger~\cite{Muller2015} focus on the city of Berlin and investigate how to predict future bookings based on the historical start of bookings time series. Finally, in~\cite{boldrini2016characterising}, we have presented an analysis of station-based car sharing in a single city. The analysis in~\cite{boldrini2016characterising} is more oriented to issues related to the presence of stations (their capacity, how their behaviour can be mathematically modelled using queueing theory, etc.) and suffers from the lack of vehicle identifiers in the studied dataset. The technique used in~\cite{boldrini2016characterising} for detecting station usage is adapted here to the free floating case, but the analysis presented here is richer, because richer is the dataset extracted from the free floating car sharing operator.

Finally, car sharing data have recently also been analysed from the perspective of data visualization\footnote{\url{http://labs.densitydesign.org/carsharing/}}, which is essential to make transportation data usable and ``communicable'' to stakeholders.

\subsection{Knowledge mining for other transportation systems}

From the methodology standpoint, this work is close to \cite{o2014mining,sarkar2015comparing}, in which bike-sharing, rather than car-sharing, systems have been analysed. Due to the different nature of the two systems, people use them differently, hence the results obtained for bike sharing systems cannot be applied directly to car sharing. However, similar methodologies can be exploited, e.g., to group stations based on how they are used by the customers.

%[C'e' tanto su bike sharing, anche perchŽ spesso sono sistemi semi-pubblici, mentre con il car sharing e' un po' diverso. \cite{austwick2013structure,o2014mining,sarkar2015comparing}]

This work is also orthogonal to the research efforts in the area of car pooling/ride sharing~\cite{trasarti2011mining,santi2014quantifying}. The idea of car pooling/ride sharing is that people may share a vehicle (be it a private or public vehicle, e.g., a taxi cab) to perform their trips.
%Both car sharing and ride sharing have the net effect of reducing the number of circulating cars. However, with car pooling users need to coordinate with each others, which makes it more suitable for longer distances than for shorter ones. In addition, due to the ongoing electrification of vehicles, the cars of the future are expected to be smaller in size~\cite{mitchell2010reinventing}, hence there will be less room for sharing a ride than for sharing the car itself. At present, however, carpooling represent an effective strategies for reducing the number of circulating cars. 
Works in the area of car pooling typically focus on the amount of rides that can be shared, based on the historical or real-time trajectories of users, hence their focus is very different from that of this work. 
%For example, in~\cite{trasarti2011mining} Trasarti~et~al. analyse a dataset of GPS observations of 2,107 real car users in Tuscany (Italy) in a time period of 12 days, and extract the percentages of users whose trajectories match (hence would be good candidates for sharing a ride). In~\cite{santi2014quantifying} instead, taxi trips performed in New York city in 2011 are mined, and a shareability network is extracted that quantifies trade-offs between the collective benefits of sharing a taxi ride and passenger discomfort.

\subsection{Operations models for car sharing}

\begin{figure}[htbp]
\begin{center}
\includegraphics[scale = 0.5]{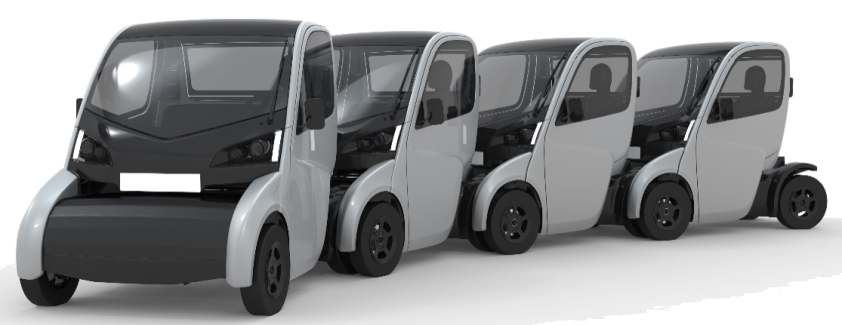}
\caption{The ESPRIT car.}
\label{fig:esprit}
\end{center}
\vspace{-20pt}
\end{figure}

As one of the pillars of a smart transportation system, car sharing has recently been the subject of much research. The research activity on this area has focused both on short and long term strategic decisions. The latter involve problems like planning the station/parking infrastructure~\cite{correia2012optimization,boyaci2015optimization,biondi2016optimal_planning} or planning the recharging infrastructure. The former is focused on decisions such as when and how to redistribute shared vehicles~\cite{ijrr12_fluid,Kek2009,Febbraro2012} or when and how to recharge them~\cite{rottondi2014complexity,biondi2016optimal_charging}. Some researcher have gone for a radically new approach to car sharing, based on the idea that some of the above problems cannot be efficiently solved  without changing the underlying car structure. This is the case of the H2020 ESPRIT project\footnote{\url{http://www.esprit-transport-system.eu/}}, whose goal is to develop purpose-built, lightweight L category electric vehicles that can be stacked together (Figure~\ref{fig:esprit}) for an efficient redistribution of fleets and a smartly-balanced and cost efficient transport system.

To address the above problems, optimisation frameworks and operational decision tools for car sharing systems have been studied in the literature, but the proposed solutions have often been evaluated either on simulated scenarios~\cite{nourinejad2015vehicle,uesugi2007optimization} or using as input the demand (in terms of origin/destination matrix) obtained from surveys~\cite{jorge2014comparing,correia2012optimization}. On the contrary, the availability of a statistical characterisation of the general properties of real car-sharing systems, as well as a precise understanding of their emerging trends, is essential to both researchers and operators in order to design more effective decision support tools, and for the calibration and validation of simulations of car sharing systems. Thus, data-driven analysis as those presented in this paper can be exploited to both  drive and evaluate solutions for the supply-side of car sharing.

%=================================================================================
\section{The dataset}
\label{sec:dataset}
\noindent
The dataset comprises pickup and drop-off times of vehicles in 10 European cities for one of the major free-floating car sharing operator. For nine of these cities, data has been collected between May 17, 2015 and June 30, 2015. For the tenth city, data covers the period from March 11, 2016 to May 12, 2016. The data has been collected every $1$ minute using the available public API, which yields responses in the form of JSON files. Errors in the data collection process are due to technical problems on the booking website, in which cases corrupted entries have been discarded from the dataset. Each entry in the dataset describe the lon-lat position of available shared vehicles in the car sharing system, plus additional information such as fuel/battery level, cleaning level, and so on. Due to faulty GPS systems, the reported coordinates may be inaccurate. For this reason the dataset has been preprocessed and coordinates that are manifestly invalid (e.g., cars available in different countries) have been discarded. Data preprocessing and analysis has been carried out in R. 

\begin{table}[htp]
\caption{Summary of dataset}
\begin{center}
\begin{tabular}{l|r|r|l|l|cl}
City & Trips & Vehicles & Date trace starts & Date trace ends & Trace duration\\
\hline \hline
City\#1 & 49901 & 349 & 2015-05-17 00:00:00 & 2015-06-30 23:58:22 & 45 days\\
\hline
City\#2 & 223044 & 981 & 2015-05-17 00:00:00 & 2015-06-30 23:59:25 & 45 days\\
\hline
City\#3 & 18944 & 198 & 2015-05-17 00:00:01 & 2015-06-30 23:58:24 & 45 days\\
\hline
City\#4 & 12168 & 194 & 2015-05-17 00:00:01 & 2015-06-30 23:54:12 & 45 days\\
\hline
City\#5 & 156080 & 686 & 2015-05-17 00:00:01 & 2015-06-30 23:59:25 & 45 days\\
\hline
City\#6 & 81862 & 499 & 2016-03-11 00:00:00 & 2016-05-12 23:54:31 & 63 days\\
\hline
City\#7 & 99515 & 584 & 2015-05-17 00:00:01 & 2015-06-30 23:58:24 & 45 days\\
\hline
City\#8 & 15612 & 250 & 2015-05-17 00:00:01 & 2015-06-30 23:51:03 & 45 days\\
\hline
City\#9 & 25091 & 299 & 2015-05-17 00:00:01 & 2015-06-30 23:53:10 & 45 days\\
\hline
City\#10 & 144474 & 829 & 2015-05-17 00:00:00 & 2015-06-30 23:59:25 & 45 days\\
\end{tabular}
\end{center}
\label{tab:dataset_summary}
\end{table}%

Given the nature of our dataset, movements of cars have to be inferred from their unavailability during a certain time frame. Thus, when a car disappears from location A to later reappear at location B, we assume that the car has been picked up for a trip. We have no explicit way for distinguishing between regular customer trips and maintenance trips (e.g., cars that have been picked up by the car sharing operator for cleaning or repairing), as we simply observe a car disappearing from the map. We also have no direct information about the trajectory followed by the shared vehicle between the start and end point of its trip. In order to integrate this information into the dataset, we have queried Google Maps asking for directions and expected travel time between the source and destination coordinates of each trip at its specific starting time. The result of the Google Map query is used to compute the expected distance in kilometers between the start and end point plus the expected journey time. How they are related to the actual distance travelled during the trip is discussed in Section~\ref{sec:trips}.

\begin{figure}[h]
\begin{center}
\begin{minipage}[c]{0.45\linewidth}
\includegraphics[scale=0.3]{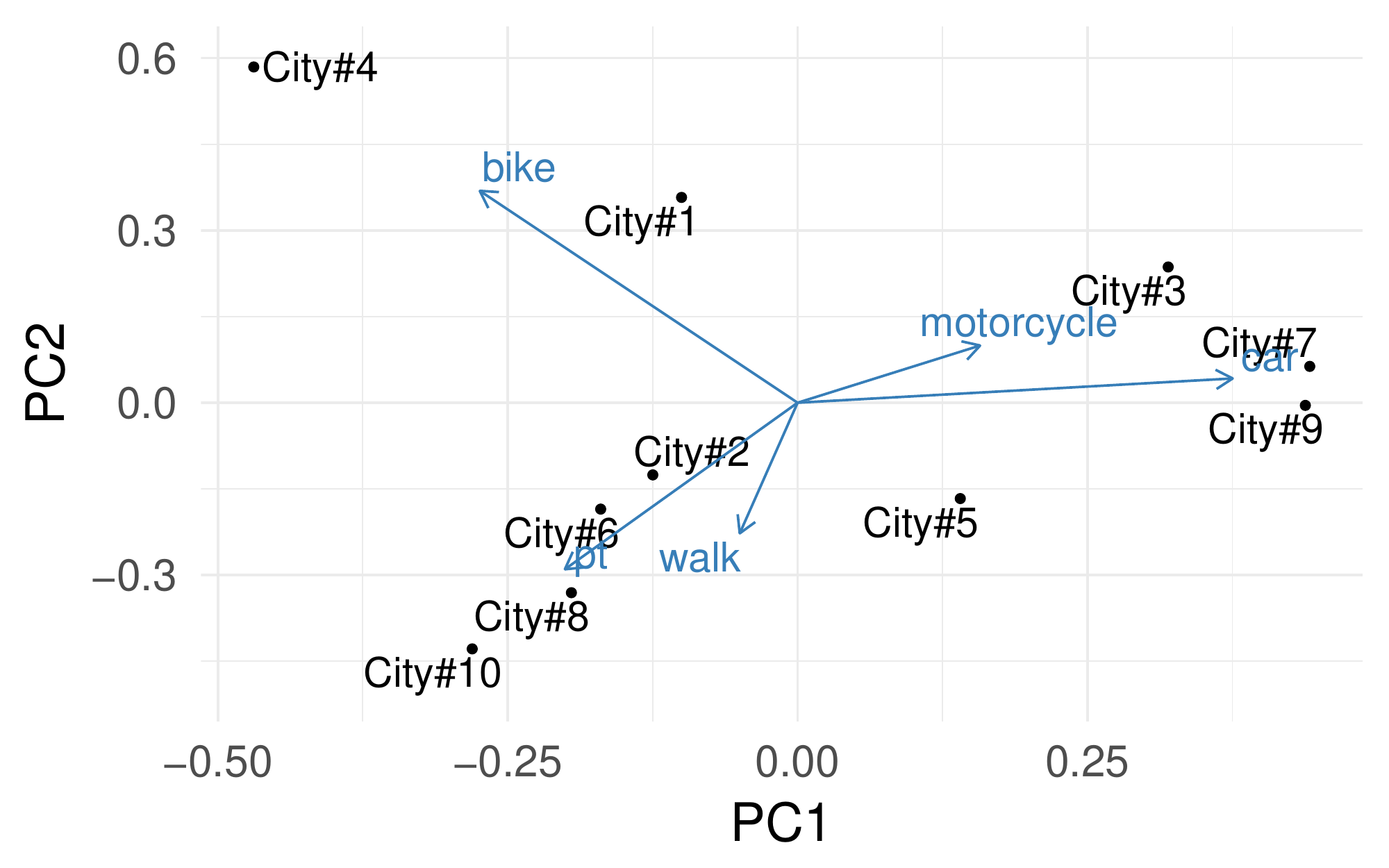}
\caption{Modal split: Principal Component Analysis}
\label{fig:modal_split_pca}
\end{minipage}
\hspace{25pt}
\begin{minipage}[c]{0.4\linewidth}
\includegraphics[scale=0.3]{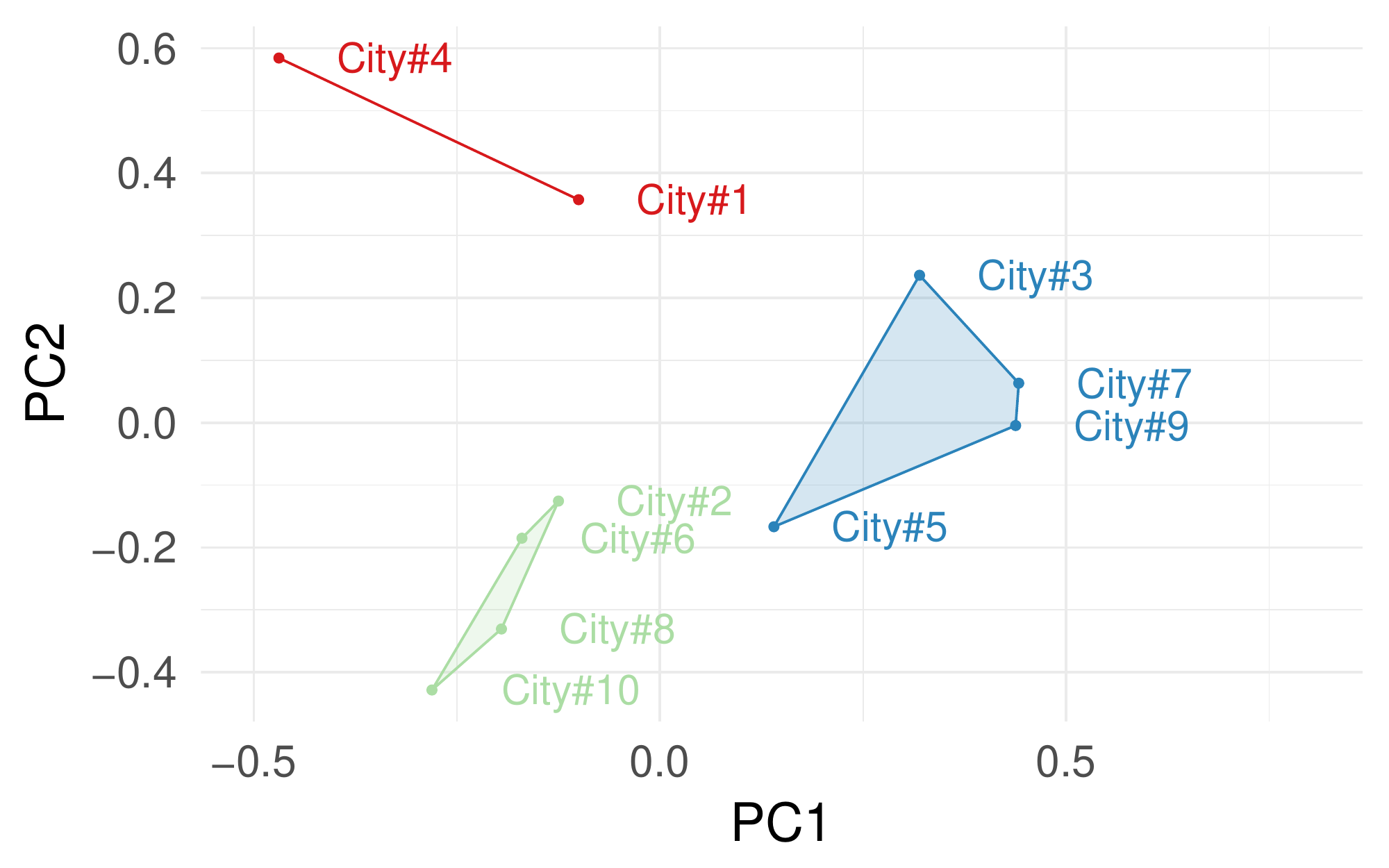}
\caption{Modal split: $k$-means clustering}
\label{fig:modal_split_clustering}
\end{minipage}
\vspace{-20pt}
\end{center}
\end{figure}

In order to understand the main characteristics of the ten cities in which the car sharing under study is operating, we have extracted information (summarised in Table~\ref{tab:cities_general_info}) from the Eurostat's City Urban Audit database\footnote{http://ec.europa.eu/eurostat/web/cities/data/database}. 
\begin{table}[htp]
\caption{General information on the 10 cities}
\begin{center}
\begin{tabular}{l|r|r|r|r|r}
City & GPD per capita & Population & Area & Population Density & Education\\
\hline \hline
City\#1 & 46952 & 853312 & 165.76 & 5147.876 & 254000\\
\hline
City\#2 & 35627 & 3520031 & 891.68 & 3947.639 & 752300\\
\hline
City\#3 & 31547 & 382929 & 102.32 & 3742.465 & 59627\\
\hline
City\#4 & 70183 & 559440 & 88.25 & 6339.263 & 152817\\
\hline
City\#5 & 87786 & 1368590 & 181.67 & 7533.380 & 224256\\
\hline
City\#6 & 46377 & 1450381 & 310.70 & 4668.107 & 371200\\
\hline
City\#7 & 55385 & 2874529 & 1287.36 & 2232.887 & 415766\\
\hline
City\#8 & 81395 & 939238 & 187.16 & 5018.369 & 234787\\
\hline
City\#9 & 74725 & 886837 & 130.17 & 6812.910 & 109314\\
\hline
City\#10 & 58140 & 1867582 & 414.87 & 4501.608 & 232009\\
\end{tabular}
\end{center}
\label{tab:cities_general_info}
\end{table}%
From the same database, we have also extracted information about the modal split in each city. Figure~\ref{fig:modal_split_pca} summarises the main transportation mode in each city as resulting from the PCA applied to the reported modal share. We can identify three main classes of cities: one in which motorised modes dominate, one in which public transport (and hence walking) are more important, and one in which people move prevalently by bike. This grouping is confirmed by Figure~\ref{fig:modal_split_clustering}, which shows the results of $k$-means clustering applied to the ten cities (the optimal number of clusters, 3, is obtained using the within-sum-of-squares method). 

An interesting feature of this dataset is that it contains entries for two cities (\copenhagen{} and \stockholm{} in our analysis) for which the car sharing operator has now shut down service. In the rest of the paper, when relevant, we will correlate this information with what is observed in these cities in terms of usage.

%=================================================================================
\section{The vehicle perspective}
\label{sec:analysis_vehicles}
\begin{figure}[h]
\begin{center}
\begin{minipage}[c]{0.45\linewidth}
\includegraphics[scale=0.35]{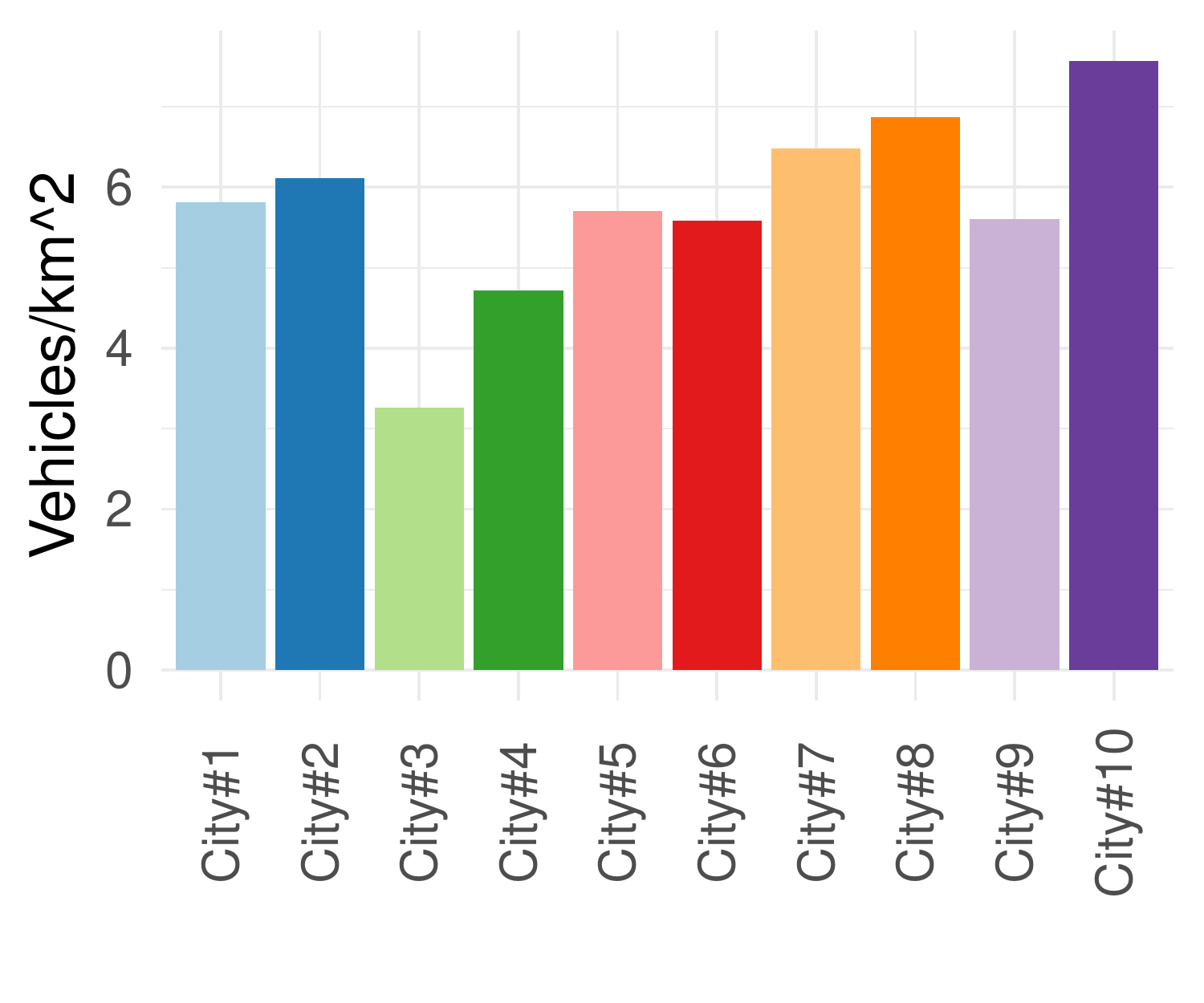}
\caption{The number of vehicles per squared km of operational area.}
\label{fig:vehicles_per_km2}
\end{minipage}
\hspace{25pt}
\begin{minipage}[c]{0.4\linewidth}
\includegraphics[scale=0.35]{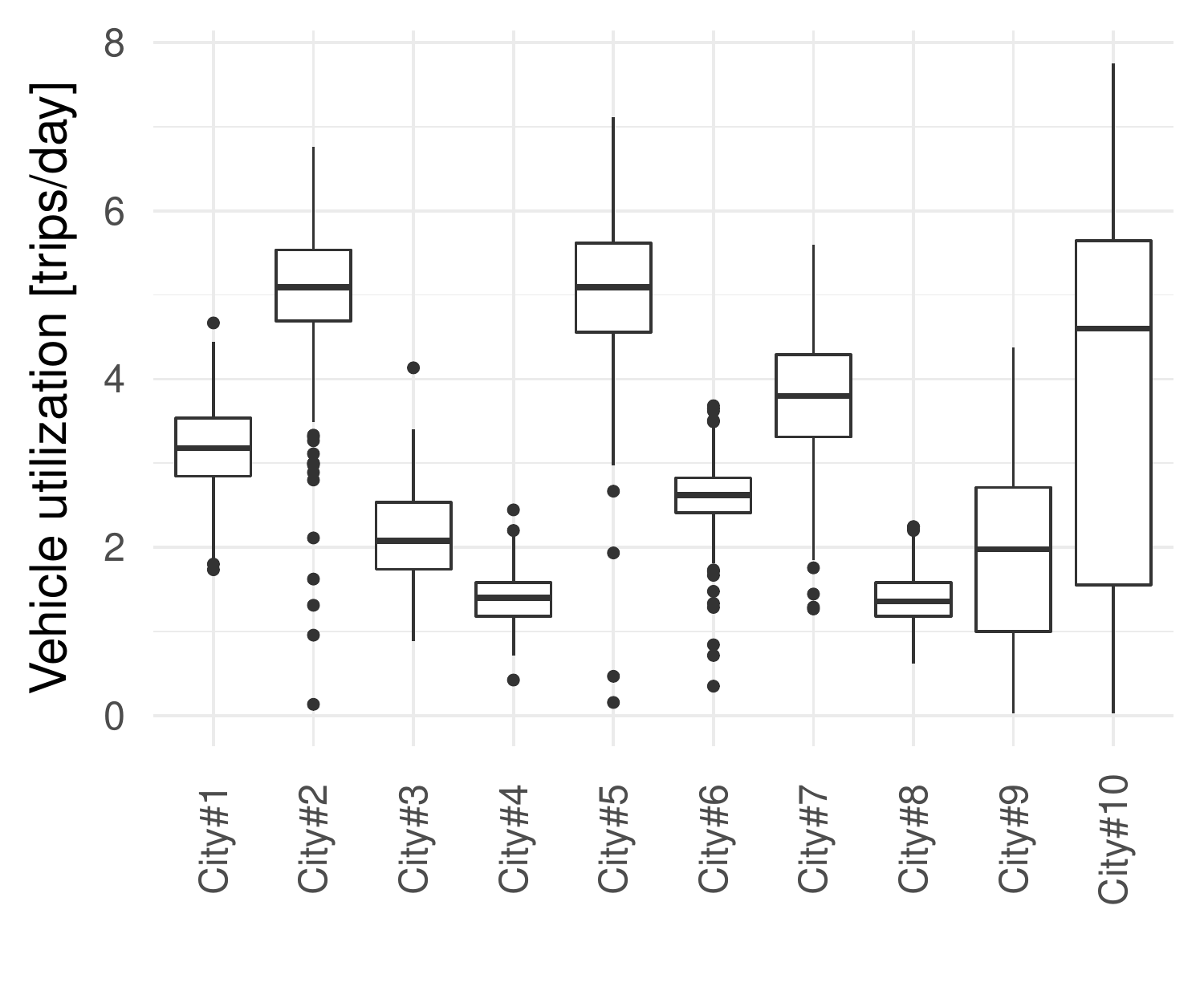}
\caption{Utilisation rate for vehicles in the shared fleet.}
\label{fig:vehicle_utilisation}
\end{minipage}
\vspace{-25pt}
\end{center}
\end{figure}

\noindent
In this section we study our dataset from the perspective of vehicles. Specifically, we want to understand if vehicles are used differently in the different cities of our dataset. First of all, we analyse if cities are unbalanced in terms of number of vehicles per squared kilometres of operational area. Figure~\ref{fig:vehicles_per_km2} shows, that in most cities, about six vehicles per squared kilometre are deployed. The most relevant exceptions are \firenze{} and \copenhagen{}, which have a smaller fleet density than the average one, and \wien{} with a higher fleet density than the average one. As better explained in the following, a small fleet density is not a sufficient condition for low profitability of the car-sharing service. Indeed, the service has been shut down in \copenhagen{} (and not in \firenze{}, which features a similar density) and in \stockholm{} (which is not even in the lowest range of density).

An index that is often used as a measure of car sharing success is the \emph{vehicle utilisation rate}, defined as the number of daily trips per vehicle. A higher value means that vehicles are used intensively in the city, hence the car sharing service is more profitable. Please note that long trips in which customers keep the shared vehicle for a long time are not the target of car sharing services but belong to the class of long-term rental. For this reason, the vehicle utilisation rate, with its ability to capture the short and frequent trips, is a direct measure of car sharing effectiveness. Figure~\ref{fig:vehicle_utilisation} shows the utilisation rate in the ten cities. It is clear how vehicles in some cities are much more utilised than in others, even 2-3 times more. It is also interesting to note that the vehicle utilisation rate is the lowest in the two cities (\copenhagen{} and \stockholm{}) where the service has been shut down months after we had collected this dataset. 

\begin{figure}[h]
\begin{center}
\includegraphics[scale=0.5]{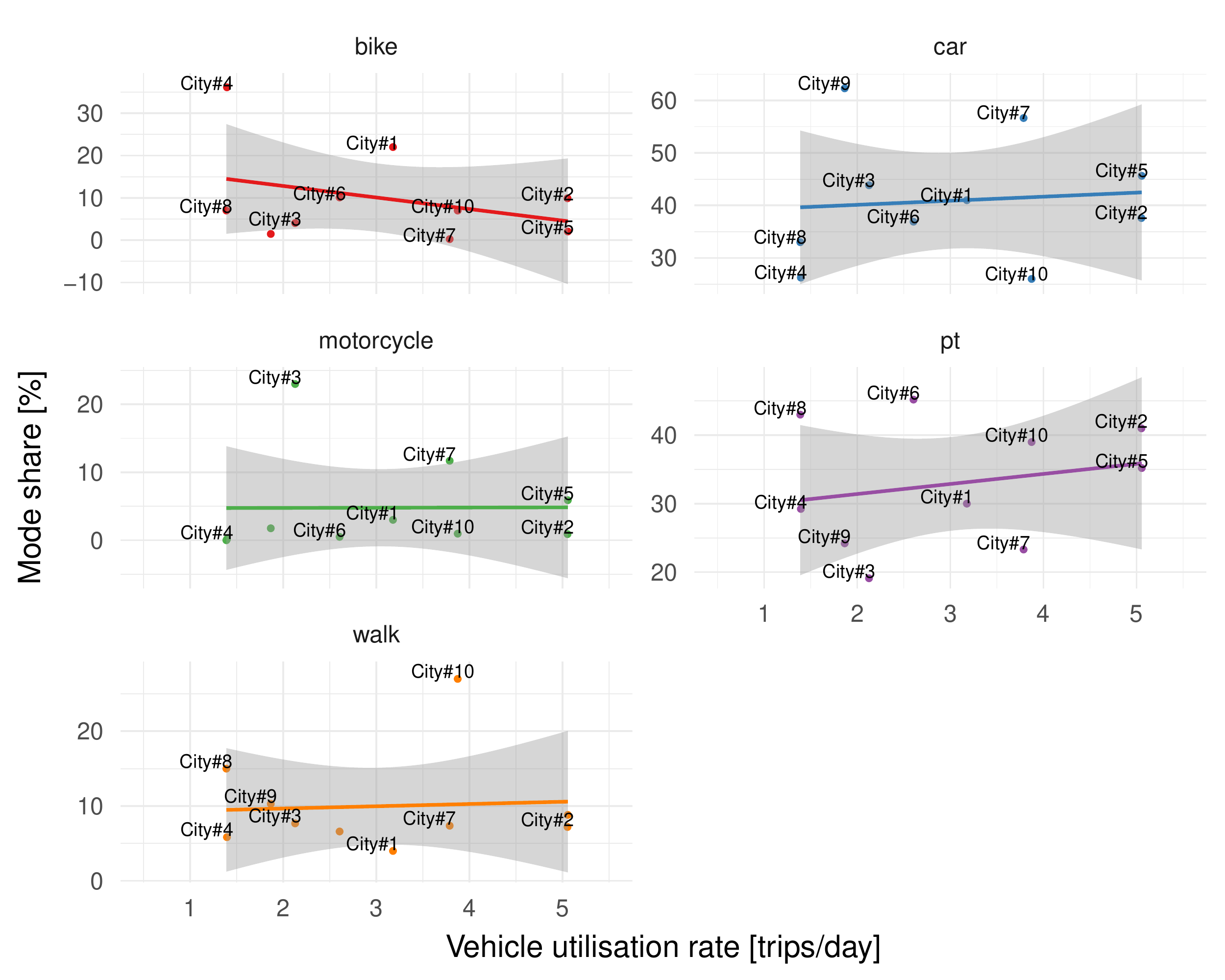}
\caption{Vehicle utilisation VS modal split.}
\label{fig:utilisation_vs_modal_split}
\end{center}
\end{figure}

While from Figure~\ref{fig:vehicle_utilisation} it is clear that there are underperforming and overperforming cities, it is yet not clear what characteristics of these cities may be affecting this performance. Given the clear differences in terms of modal split in the ten cities, with the three clusters of mode share discussed in Section~\ref{sec:dataset}, in the next set of plots we study whether a correlation between modal split and car sharing performance may exist. To this aim, in Figure~\ref{fig:utilisation_vs_modal_split} we plot the vehicle utilisation against the different modes in the cities. Predominance of bike journeys correlates negatively with the car sharing performance (Pearson $r= -0.34$), while diffusion of public transport correlates positively with car sharing  ($r=0.22$). Walking and motorcycle do not correlate with car sharing usage ($r=0.06$ and $r=0.0051$, respectively), and similarly for the use of cars ($r=0.09$).

%\textcolor{blue}{So, in \copenhagen{}, bike journeys are predominant, shared cars are not used much, not even considering that the fleet density is small.}

\begin{figure}[t]
\begin{center}
\includegraphics[scale=0.4]{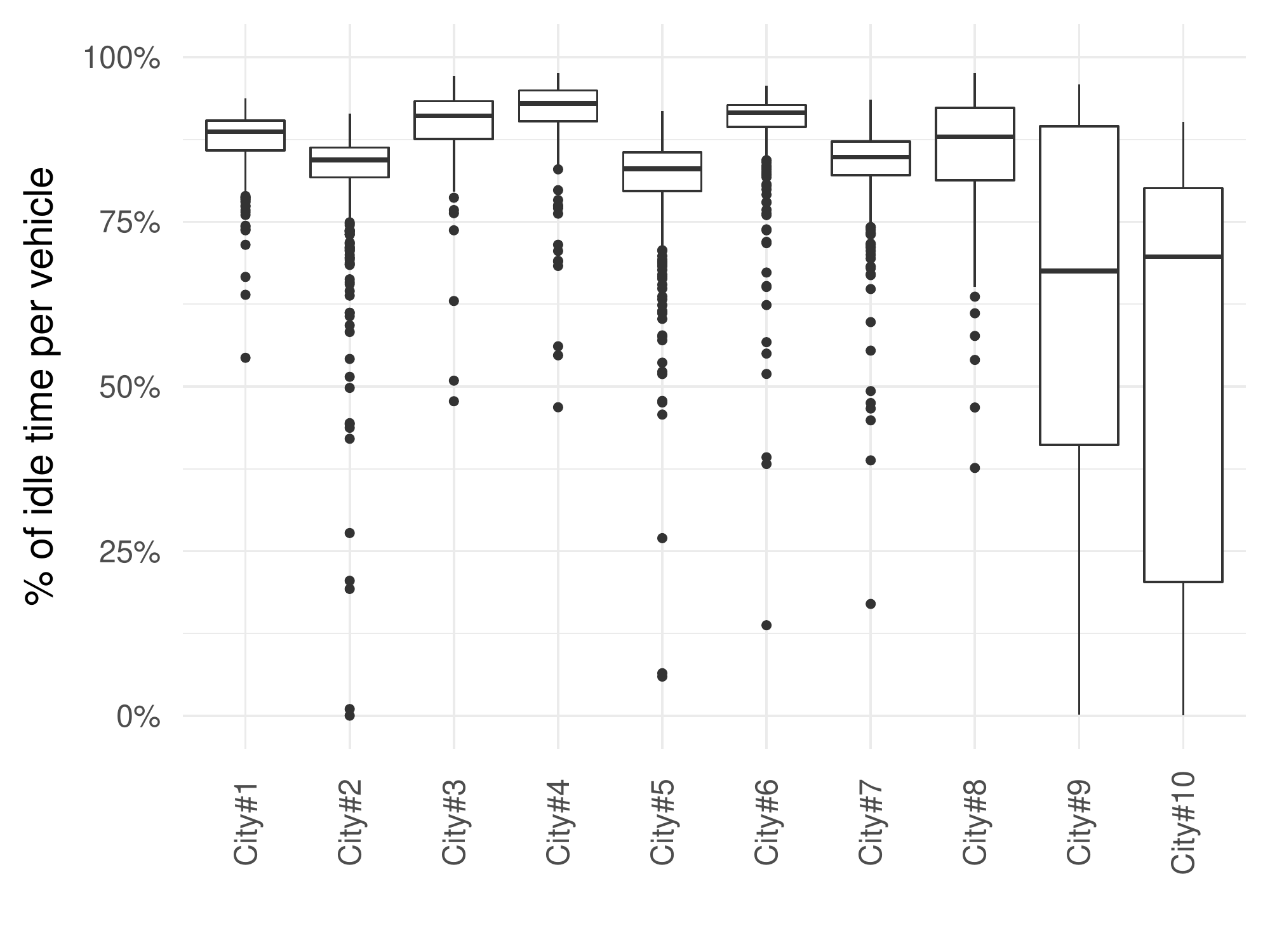} \vspace{-20pt}
\caption{Vehicle idle time.}
\label{fig:vehicle_idle_time}
\end{center}
\vspace{-20pt}
\end{figure}

Another interesting parameter that characterises vehicle usage is the vehicle idle time, i.e., what is the fraction of time the vehicle remains idly parked in the operational area. We show this quantity in Figure~\ref{fig:vehicle_idle_time}. Even in the best case, vehicles remain parked most of the time. This is good news for the research on vehicle redistribution: the operator can indeed exploit a large number of vehicles that are not used most of the time. One of the reasons they are not used is possibly that they are not where they are needed most (so-called hot spot areas). A continual fleet redistribution could address this issue and improve the overall availability experienced by customers.

%=================================================================================
\section{The trips perspective}
\label{sec:trips}
\noindent
In this section we move one step ahead, changing the perspective from vehicles to trips. As discussed in Section~\ref{sec:dataset}, we can only acquire indirect knowledge about car sharing trip based on the fact that vehicles ``disappear" from the map when they are being used by customers. The main problem is thus that we have no direct way for distinguishing regular customer trips from maintenance trips. To the best of our knowledge, the free floating car sharing system covered by the dataset was not implementing vehicle redistribution at the time the traces were collected. For this reason, we expect maintenance trips to have a signature quite different from all other trips (e.g., be longer and/or bring back the vehicle with a full battery/tank). However, since this kind of trips are also expected to be significantly fewer than regular trips, for the purpose of this analysis we will simply ignore their presence.
 
\begin{figure}[t]
\begin{center}
\begin{minipage}[c]{0.5\linewidth}
\includegraphics[scale=0.3]{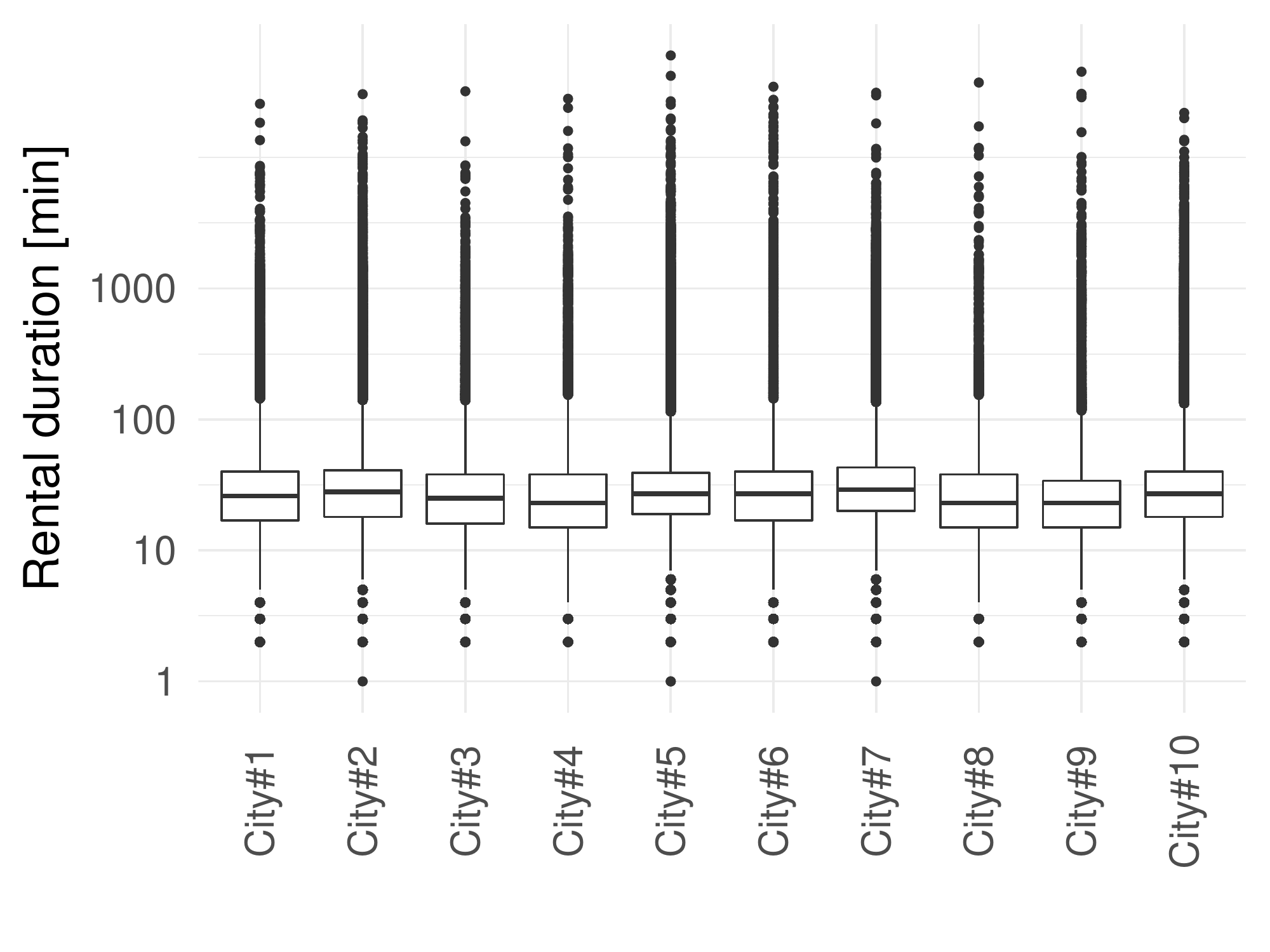}
\caption{Boxplot of rental duration.}
\label{fig:rental_duration}
\end{minipage}
\hspace{5pt}
\begin{minipage}[c]{0.4\linewidth}
\includegraphics[scale=0.3]{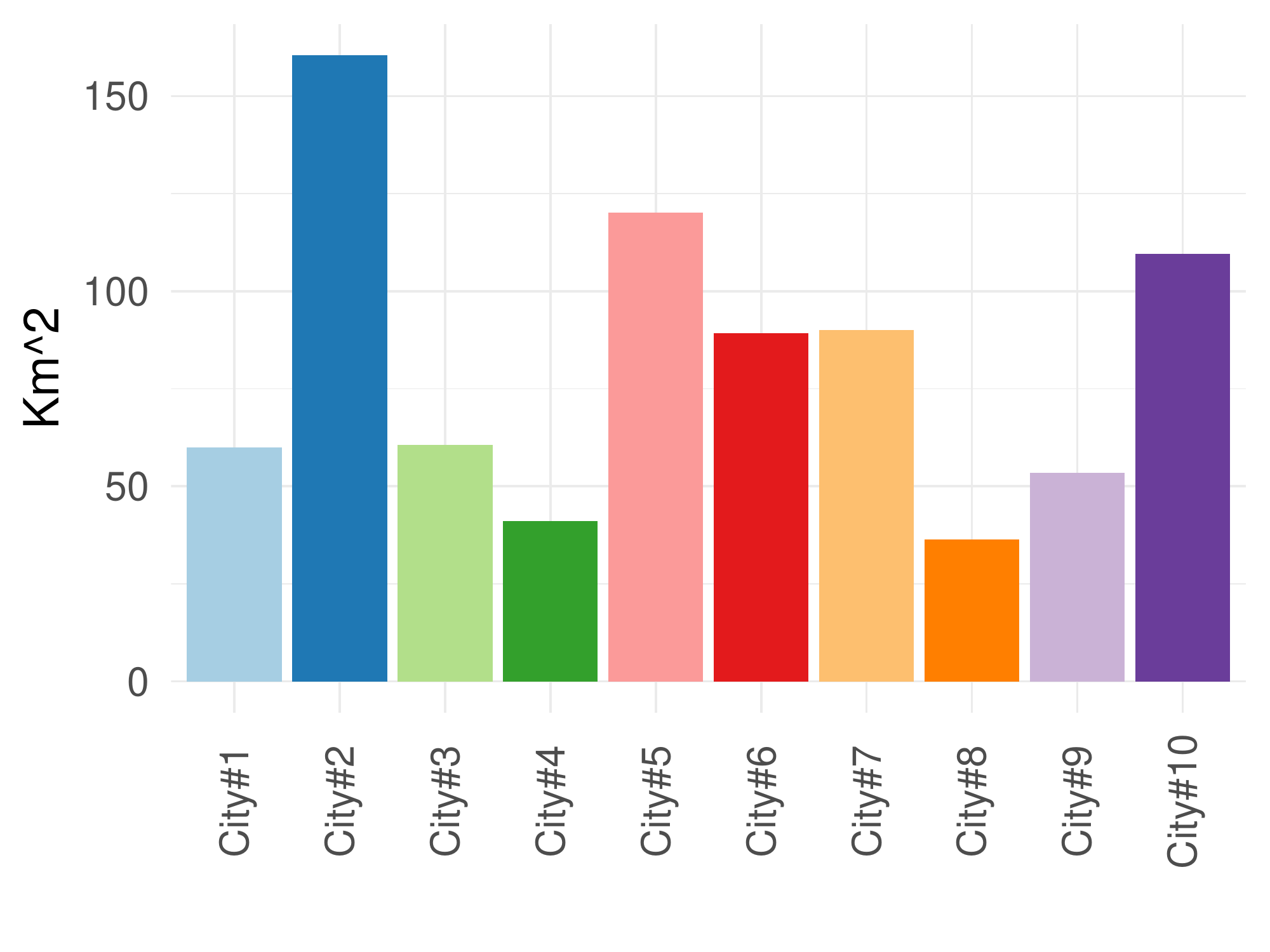}
\caption{Size of operational area.}
\label{fig:operational_areas}
\end{minipage}
\vspace{-20pt}
\end{center}
\end{figure}

We first consider the probability distribution of rental time for the trips in each city (Figure~\ref{fig:rental_duration}). The vast majority of trips last less than one hour in all cities. Moreover, all cities are basically equivalent from the rental duration standpoint, despite the fact that the size of the operational area may be significantly different (Figure~\ref{fig:operational_areas}). 

\begin{figure}[htbp]
\begin{center}
\includegraphics[scale = 0.5]{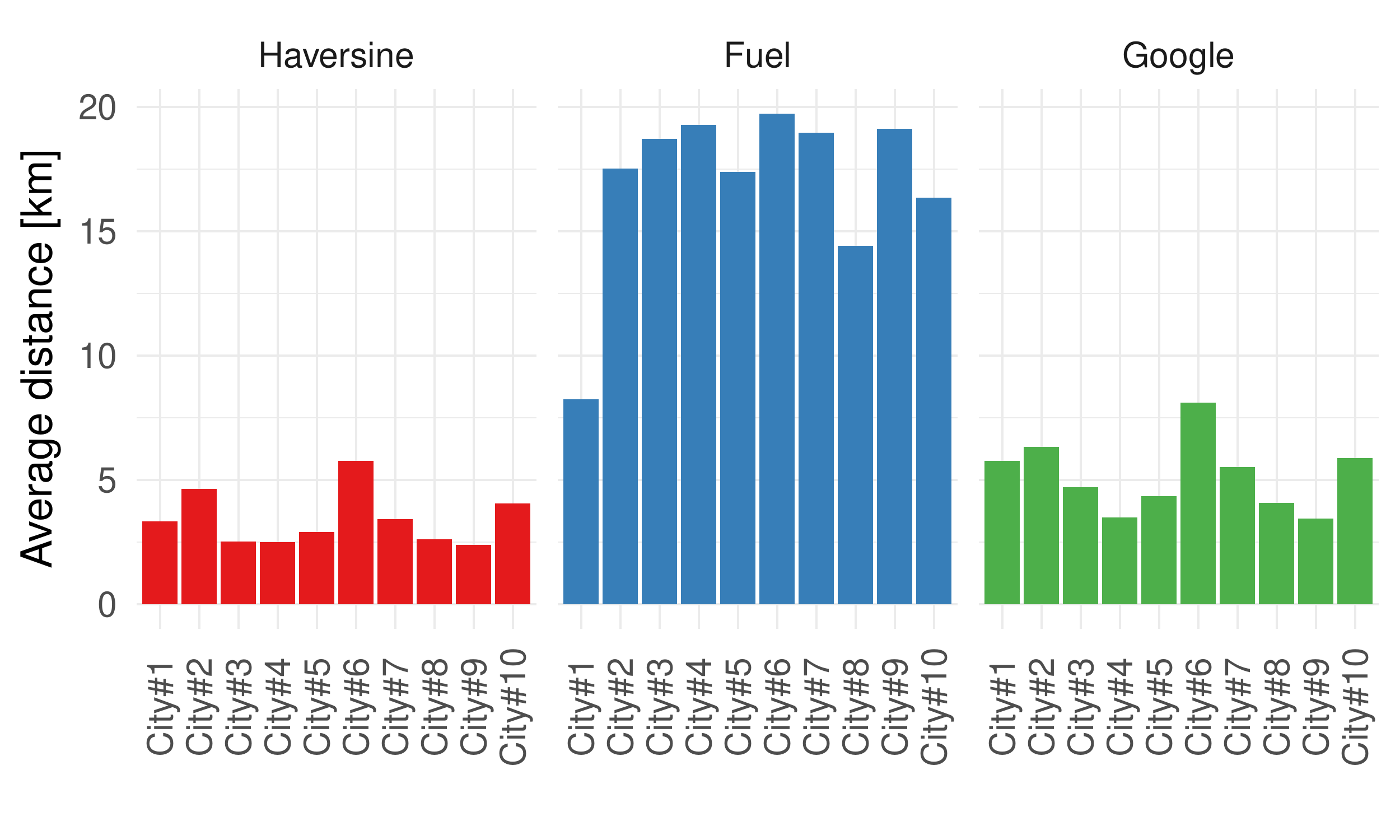} \vspace{-20pt}
\caption{Distances.}
\label{fig:distances}
\end{center}
\vspace{-20pt}
\end{figure}

Double-checking this behaviour considering the trip distance is not straightforward, since we have no direct information on the trajectory and/or intermediate stops for trips. As discussed in Section~\ref{sec:dataset}, we have tried to fill this gap with the help of the Google Maps APIs. Thus, for each rental starting at location A and ending at location B, we have collected the route recommended by Google Maps. Another indirect measure of trip distance is the battery/fuel\footnote{In just one city, the shared fleet is electric.} consumption per trip. In order to extract the distance, we have derived, for each type of vehicle in the fleet, the average battery/fuel consumption from \url{ http://www.spritmonitor.de/en/}, a website where users can track the fuel consumption of their vehicles. The average distance for the ten cities computed according to the two different methodologies discussed above is shown in Figure~\ref{fig:distances}, against the geodesic (Haversine) distance between the trip endpoints. The three measures are very loosely correlated. This is a strong indication that the start and endpoints of trips tell only a partial story about how people use shared cars for their needs.

In order to better understand the nature of trips in the free floating system covered by our dataset, we have designed a simple classifier to distinguish between one-way trips, one-way trips with intermediate stops and roundtrips (i.e., trips that come back at their starting point when they end). The classifier works as follows. Let us consider a vehicle disappearing from location A and reappearing at location B. We first compare the expected travel time according to Google between A and B. If there is a match with the rental duration (i.e, if $ T_{google} * (1 - \delta_{low}) < T_{rent} < T_{google} * (1 + \delta_{up})$, where $\delta_{low}$ and $\delta_{up}$ are two tolerance factors set to 0.1 and 0.2 respectively), then the trip is classified as one-way. If $T_{rent} < T_{google} * (1 + \delta_{up})$, i.e., if the rental duration is significantly longer than the expected time to travel from A to B, then we look at the exact coordinate of A and B. If $dist(A,B) < 500m$ and the rental time is much longer than what it would take to go from A to B at a speed of $50$km/h, then the trip is classified as two-way. Vice versa, if A and B are not close, the trip is classified as one-way with stops. 

\begin{figure}[htbp]
\begin{center}
\includegraphics[scale=0.45]{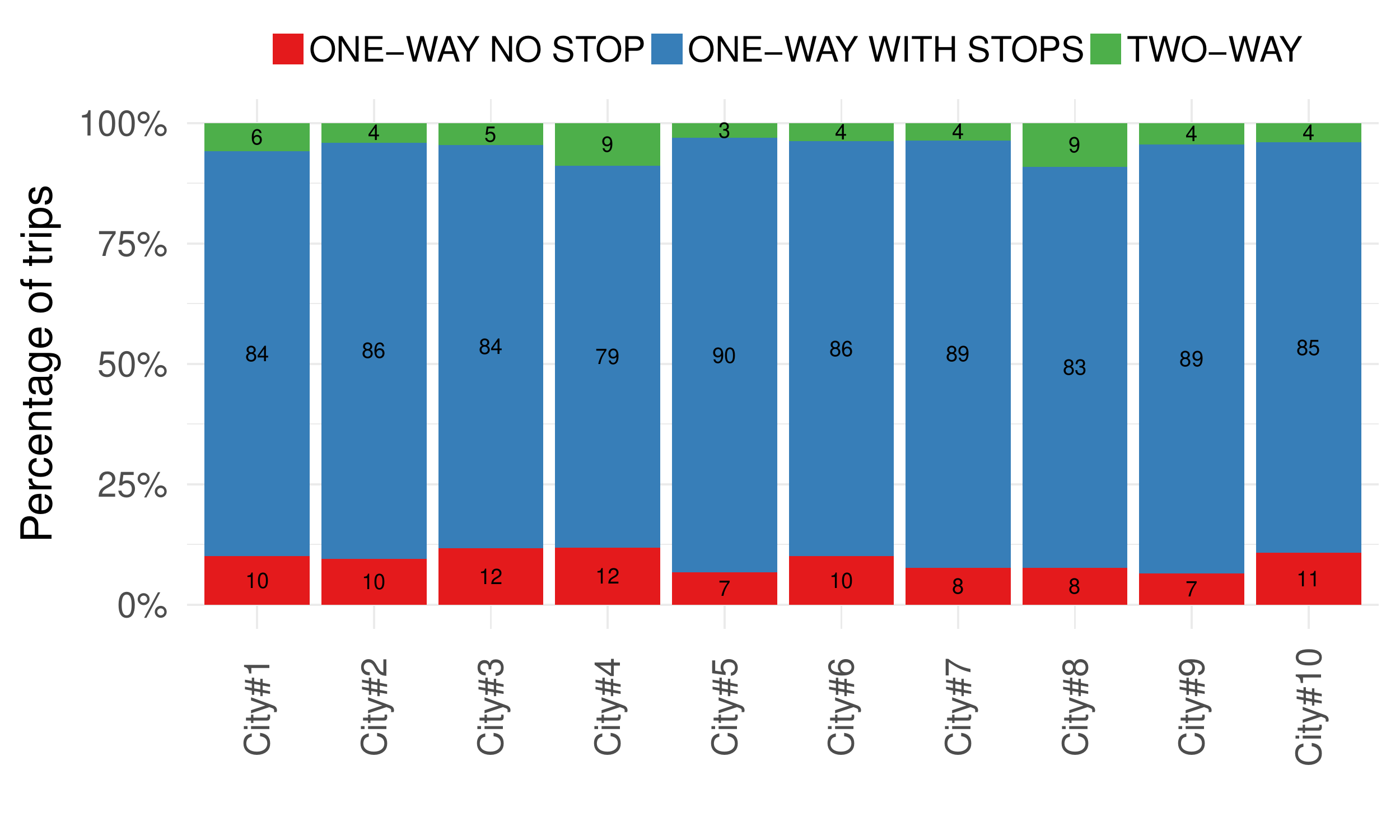}\vspace{-20pt}
\caption{Types of trips.}
\label{fig:trip_classification}
\end{center}
\vspace{-20pt}
\end{figure}

The result of the classification is shown in Figure~\ref{fig:trip_classification} (please note that we have no ground truth for verifying this result). It is interesting to observe that the vast majority of trips are one-way trips with intermediate stops. This information can shed some light on the purpose of car sharing trips. For example, car sharing trips may be more appealing to customers that have to run some errands than to commuters.

%=================================================================================
\section{The spatio-temporal perspective}
\label{sec:availability}
\noindent
This section is devoted to the characterisation of vehicle availability and system usage over time and space. In order to proceed with a spatial analysis we need to identify a meaningful spatial unit to define car availability in a given area. In fact, differently from station-based car sharing, in free floating car sharing there is no natural ``aggregation" point for vehicles, which can be freely picked up and dropped off anywhere within the operational area. We can still perform a spatial analysis of car sharing usage by dividing the operational area into smaller cells and studying what is the behaviour, over time, in each of these cells. In this work, exploiting the analogy with station-based car sharing in which station coverage varies between 200m to more than 500m~\cite{boldrini2016characterising}, we consider cells with side length 500m.

\begin{figure}[p]
\begin{center}
\includegraphics[scale=0.5]{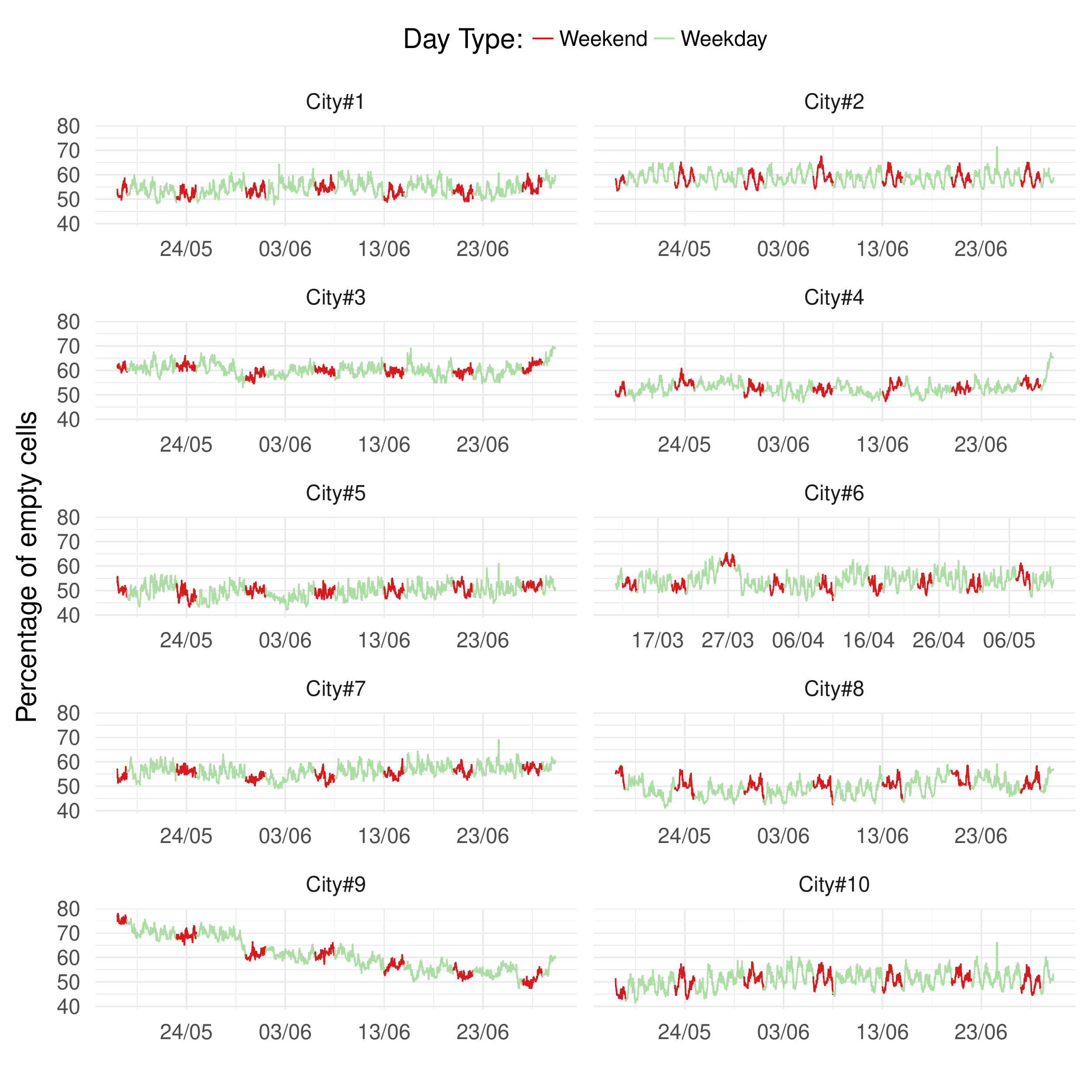}

\caption{Percentage of empty cells over time.}
\label{fig:empty_system}
\end{center}
\end{figure}

\begin{figure}[p]
\begin{center}
\includegraphics[scale=0.5]{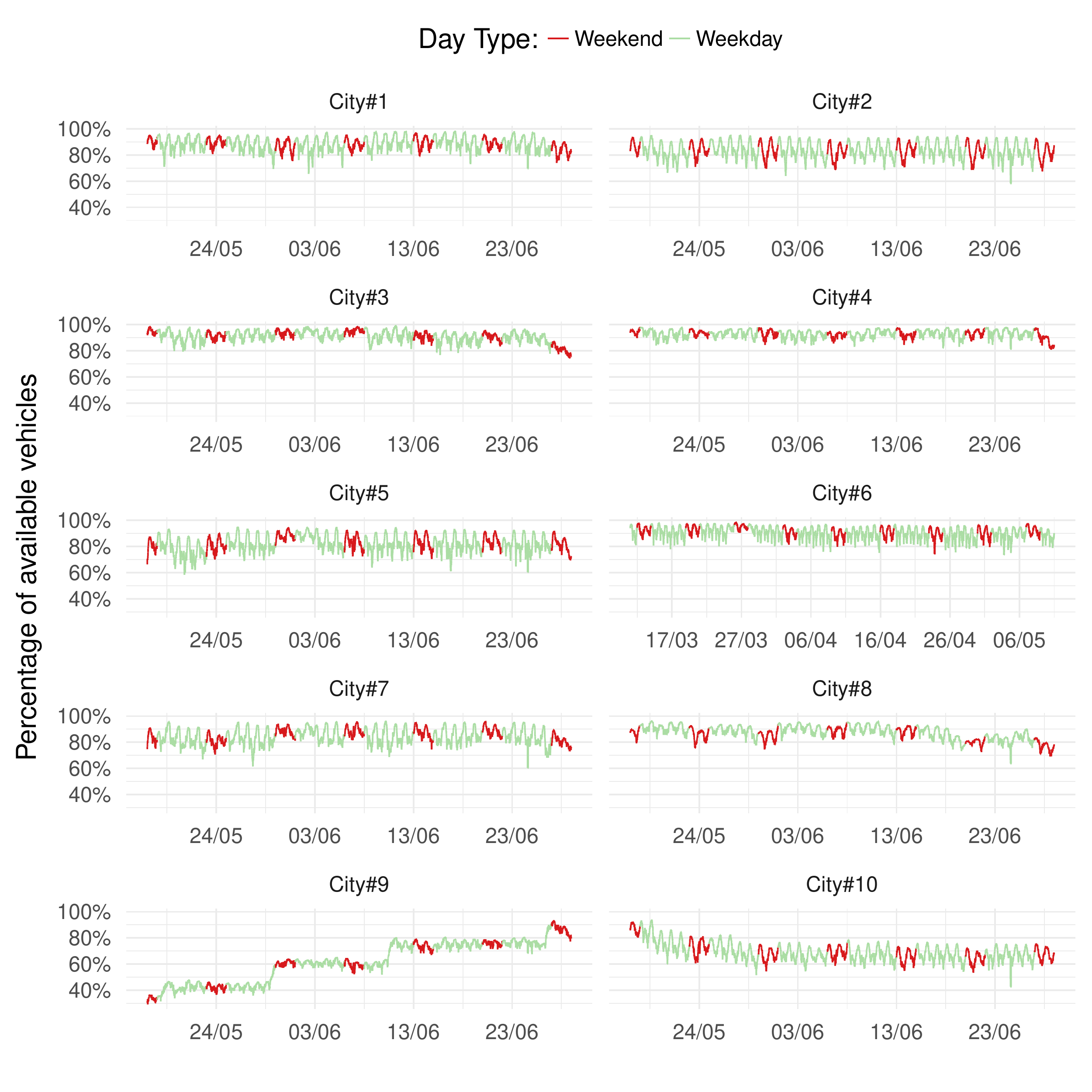}
\caption{Percentage of available vehicles over time.}
\label{fig:availability_system}
\end{center}
\end{figure}

In Figure~\ref{fig:empty_system} we show how the number of empty cells varies over time in each city. There are always a lot of cells with no vehicle, with peaks of 70\%-80\%. However, if we compare this situation with what is observed at system level in terms of vehicle availability (Figure~\ref{fig:availability_system}), it is immediate to detect a clear discrepancy: there are a lot of empty cells (Figure~\ref{fig:empty_system}) but at the same time there are also a lot of available vehicles (Figure~\ref{fig:availability_system}). This situation hints at a strong concentration of vehicles in certain areas\footnote{\torino{} represents a very special case as its car sharing system was opened just weeks before we started collecting our dataset. It seems evident that the behaviour in the city has not stabilised during the 45 days of the trace.}. In particular, it is a strong indication that macroscopic properties are not very informative when it comes to car sharing. For this reason, in the remainder of the section, we will focus on understanding the microscopic behaviour of cells in terms of their individual predictability (Section~\ref{sec:regularity}) and their similarity in usage patterns (Section~\ref{sec:cell_usage}). 

%=================================================================================
\subsection{Regularity in car sharing}
\label{sec:regularity}
\noindent
Demand predictability is one of the crucial aspects for every transportation system. In car sharing, in particular, it is of utmost importance for vehicle redistribution, whose goal is in fact to move vehicles in order to address the future demand. Clearly, there is an intrinsic limitation on the effectiveness of redistribution, which is the regularity of the demand. If the demand were totally unpredictable, redistributing vehicle would not only be useless in terms of quality of service provided to customers but also uselessly costly for the operator.

\begin{figure}[p]
\begin{center}
\includegraphics[scale=0.5]{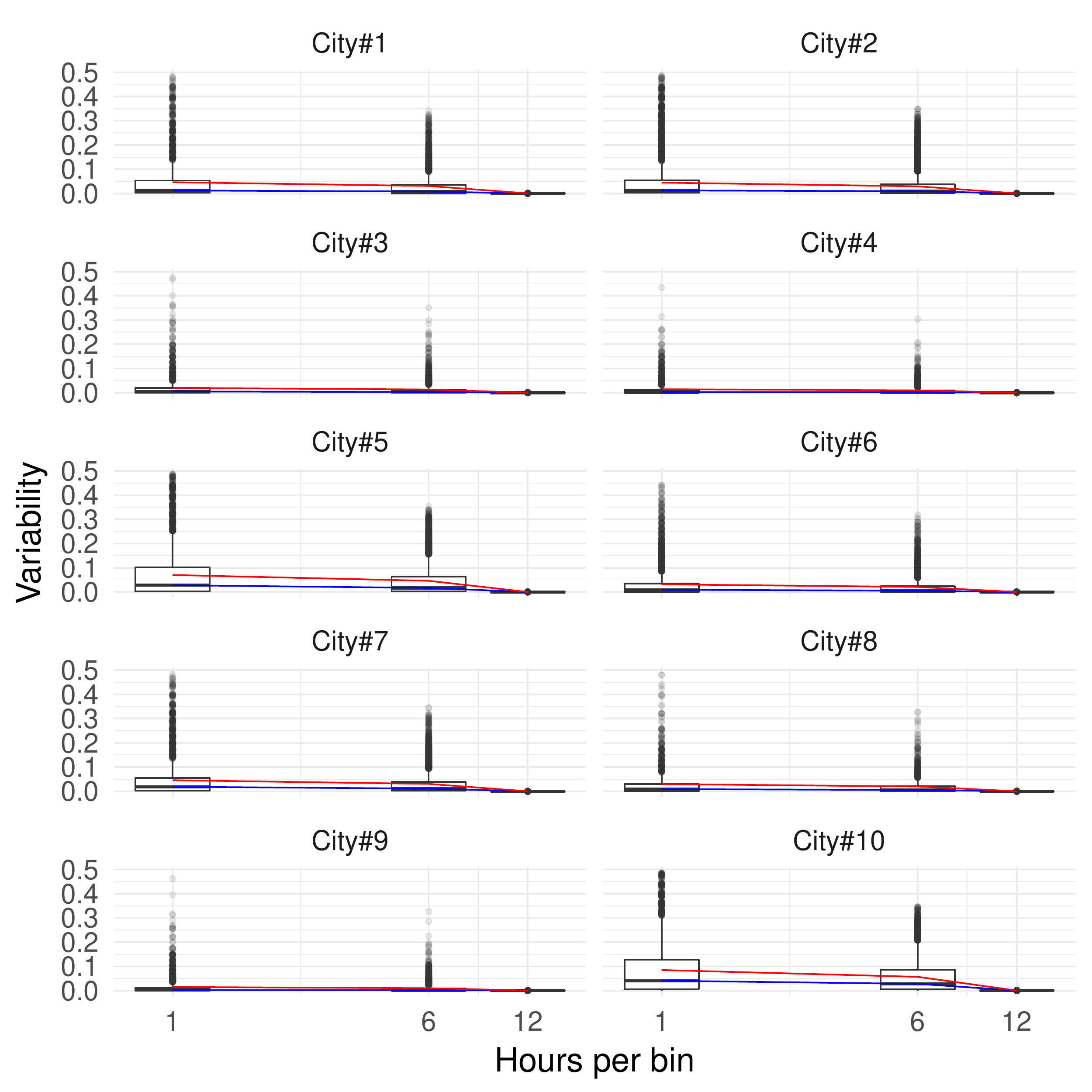}
\caption{Predictability of trips.}
\label{fig:regularity}
\end{center}
\end{figure}

We measure cell regularity in terms of the number of pickup events observed within the cell during working days. In order to measure how much the number of pickups varies across the observation period  we use the technique described by Zhong et al.~\cite{Zhong2016}. Due to space limitations, we omit the details and we refer directly to~\cite{Zhong2016} for details. The results are shown in Figure~\ref{fig:regularity}. For all cities, the vast majority of cells has an extremely predictable behaviour, with limited variability. However, the number of outliers is significant, and it should be taken into account when designing supply models for car sharing services (e.g., unpredictable cells should not be taken into account in the redistribution process).

%=================================================================================
\subsection{Cell usage patterns}
\label{sec:cell_usage}

It is expected that cells in a car sharing system are used differently by the users, but how many different usages can be identified? In order to answer this question, in the following we carry out a classification of cells based on their usage pattern. To this aim, we focus on the time series of vehicle availability in each cell and we measure how close this time series is with what we observe in other cells. We measure the time series distance using the Dynamic Time Warping (DTW) technique~\cite{esling2012time} (with Sakoe-Chiba band), then we cluster cells based on their DTW-distance using PAM clustering. For each city, the optimal number of clusters is obtained using the silhouette method. In order to be able to compare our time series, we discretise time into bins with a duration of 10 minutes. For each cell, we extract one availability value per bin by averaging the availability in the bin in different days. In addition, in order to detect variation above and below the average behaviour, we normalise the measured availability using the average availability at the cell.

\begin{figure}[h]
\begin{center}
\begin{minipage}[c]{0.5\linewidth}
\includegraphics[scale=0.3]{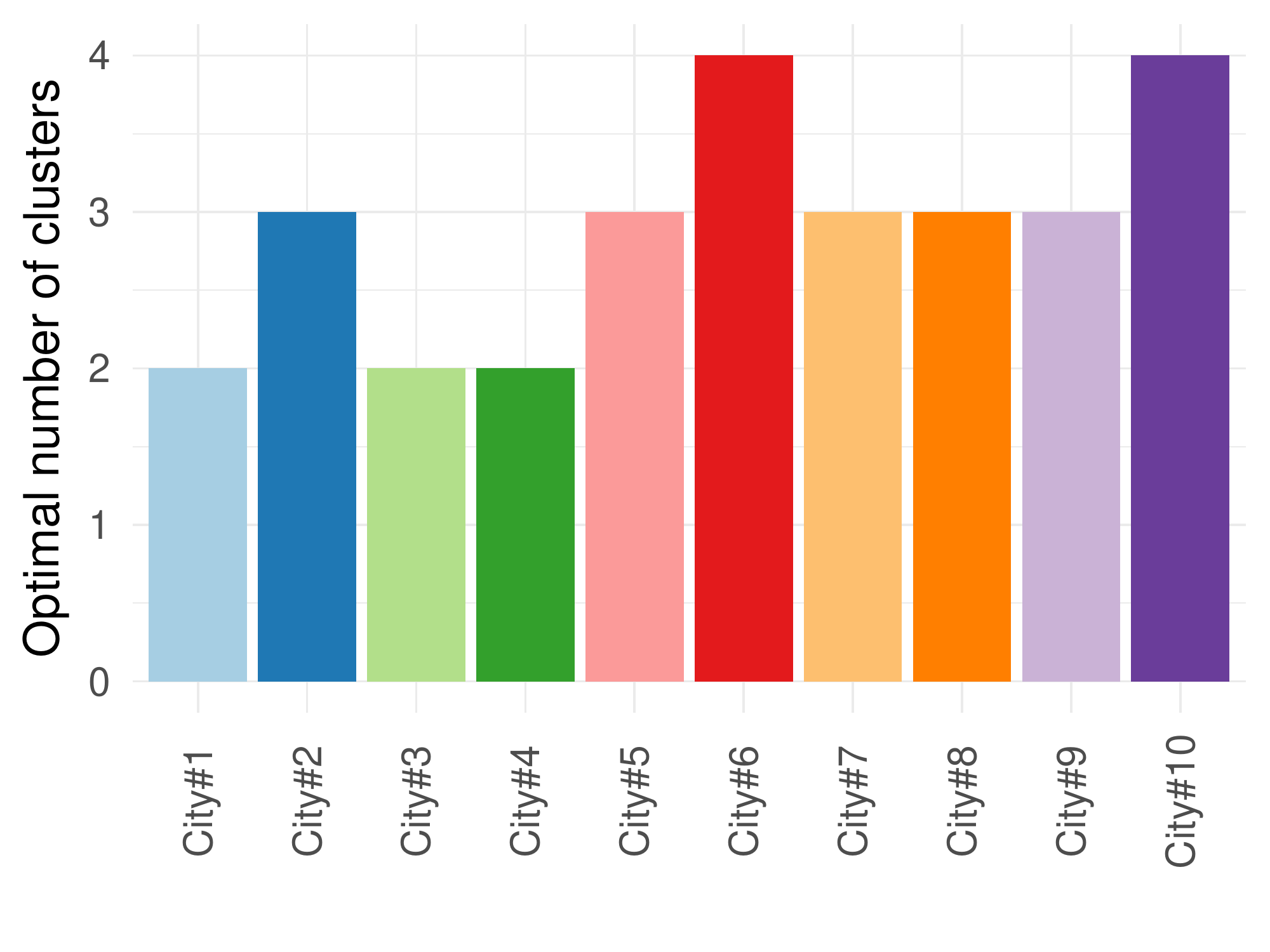}
\caption{Optimal number of clusters.}
\label{fig:opt_num_clusters}
\end{minipage}
\hspace{5pt}
\begin{minipage}[c]{0.4\linewidth}
\includegraphics[scale=0.3]{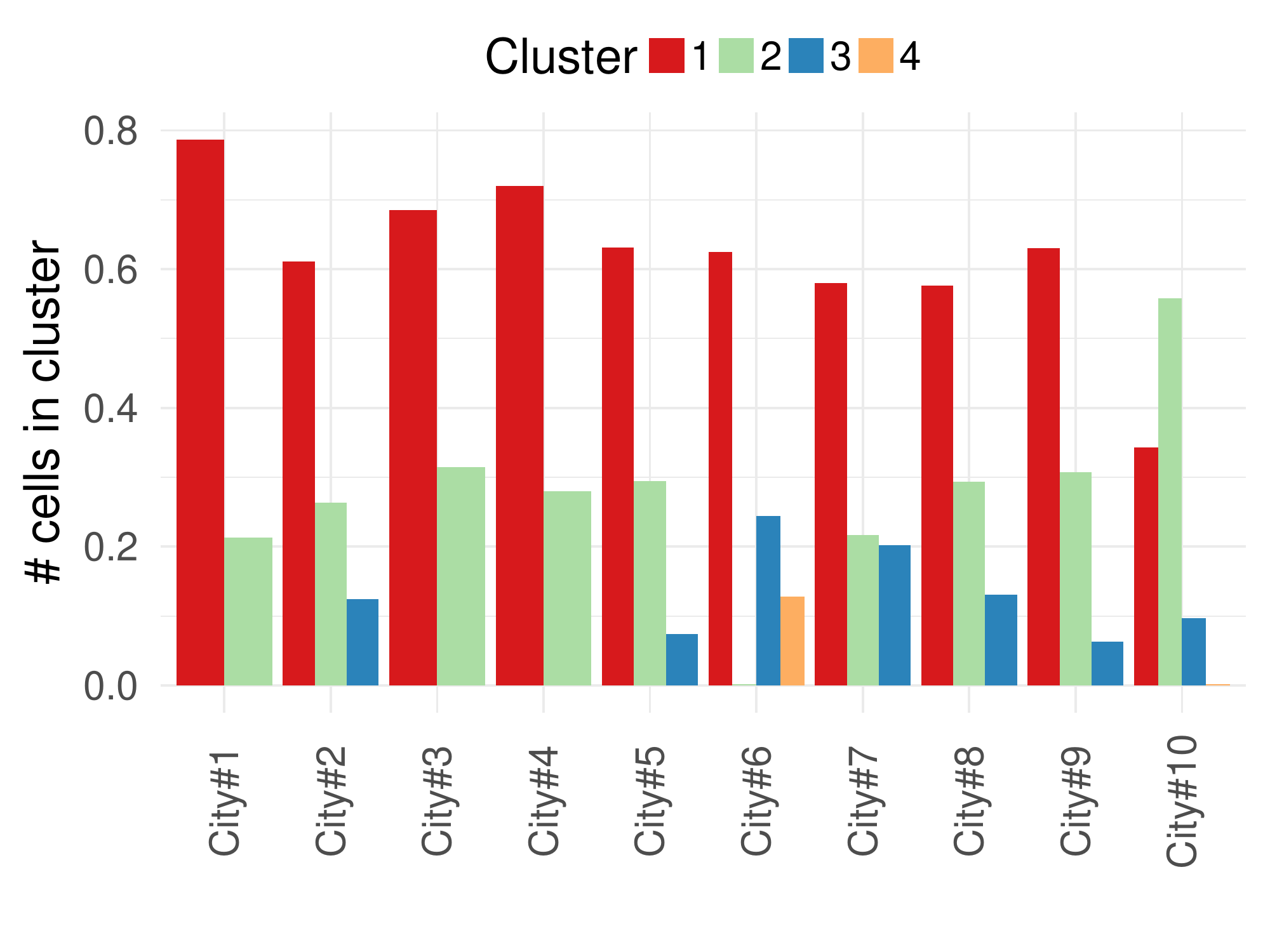}
\caption{Fraction of cells in each cluster.}
\label{fig:cells_per_cluster}
\end{minipage}%
\end{center}
\vspace{-20pt}
\end{figure}

\begin{figure}[p]
\begin{center}
\includegraphics[scale=0.5]{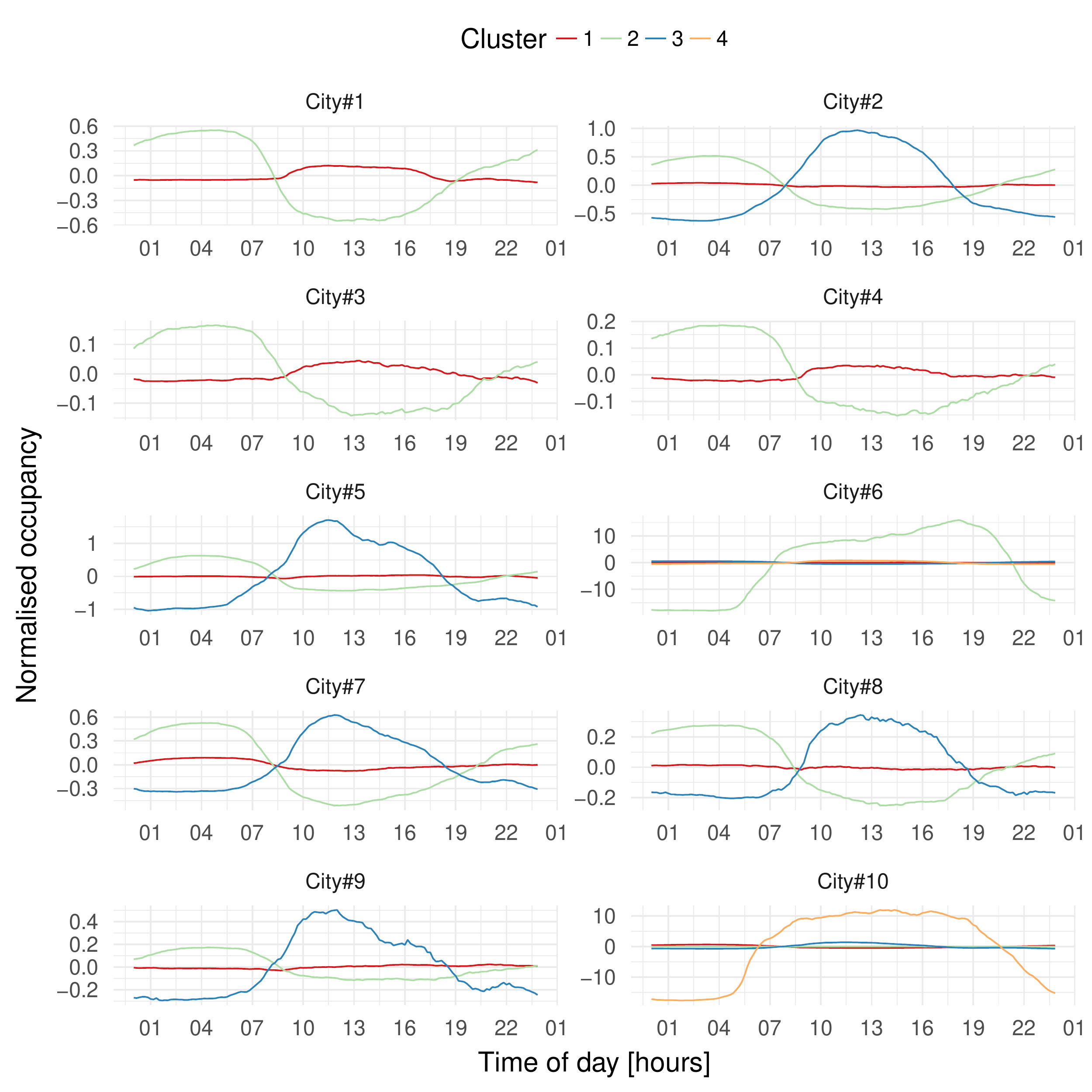}
\caption{Time series of vehicle availability per cluster in the ten cities.}
\label{fig:clusters_ts}
\end{center}
\end{figure}

The results are shown in Figures~\ref{fig:opt_num_clusters}-\ref{fig:clusters_ts}. The optimal number of clusters in all cities ranges from 2 to 4 (Figure~\ref{fig:opt_num_clusters}). However, the fourth cluster, when present, is a very special cluster, composed of just a single cell. This single cell is a very special one in the city ecosystem, and in both cities where the fourth cluster is present, this cluster comprises the airport zone. If we plot the availability time series within each cluster (Figure~\ref{fig:clusters_ts}, obtained by computing the availability in the cells belonging to the cluster), it is striking to see that the clusters highlight very characteristic cell usage. Some cells have above average availability at night and below average availability during the day. Other cells have exactly the opposite behaviour. Finally, there is a group of cells with an intermediate behaviour, where apparently no significant different in usage is detected over the whole day. It is easy to map this behaviour into the ``nature'' of the area covered by the cell: people leave residential areas in the morning and come back in the evening, while the opposite is true for commercial/business areas. As future extension of this work, we plan to compare this result again the predominant land use of each cell. Figure~\ref{fig:clusters_ts} also highlights the outlier behaviour of airport zone (which constitute the fourth cluster, when available). Airports in \muenchen{} and \wien{} see a huge variation in availability. However, the behaviour of their time series is simply a scaled version of the commercial/business pattern discussed before.

%%----------------------------------------------------------------------------------------------------------------------------------------------
%\subsection{Station capacity, car availability, parking duration}
%\label{sec:capacity_and_availability}
%%
%\noindent
%%----------------------------------------------------------------------------------------------------------------------------------------------
%\subsection{Station/cell usage classifier}
%\label{sec:clustering}
%%
%\noindent
%
%%----------------------------------------------------------------------------------------------------------------------------------------------
%\subsection{Parking time, drop-off and pickup rates}
%\label{sec:rates_and_parking_time}
%%
%\noindent
%
%%----------------------------------------------------------------------------------------------------------------------------------------------
%\subsection{Identifying problematic areas}
%\label{sec:money}
%%
%\noindent

%----------------------------------------------------------------------------------------------------------------------------------------------
\section{Identifying potential service areas}
\label{sec:service_areas}
\noindent
A critical operational aspect for car sharing operators is how to perform cleaning and maintenance. Typically, the car sharing workforce is dispatched to collect vehicles that are in need of either. However, moving workers around is expensive, and more efficient solutions could be found based on the vehicle usage in the city. As a case study, in the following we discuss how to identify potential service areas within the operational area. A potential service area is a location vehicle pass by with very high probability. A workshop could be  deployed in this area, and this would make cleaning and maintenance operations much more efficient. 

\begin{figure}[h]
\begin{center}
\subfloat[Tolerance window: 30 days]{\includegraphics[scale=0.4]{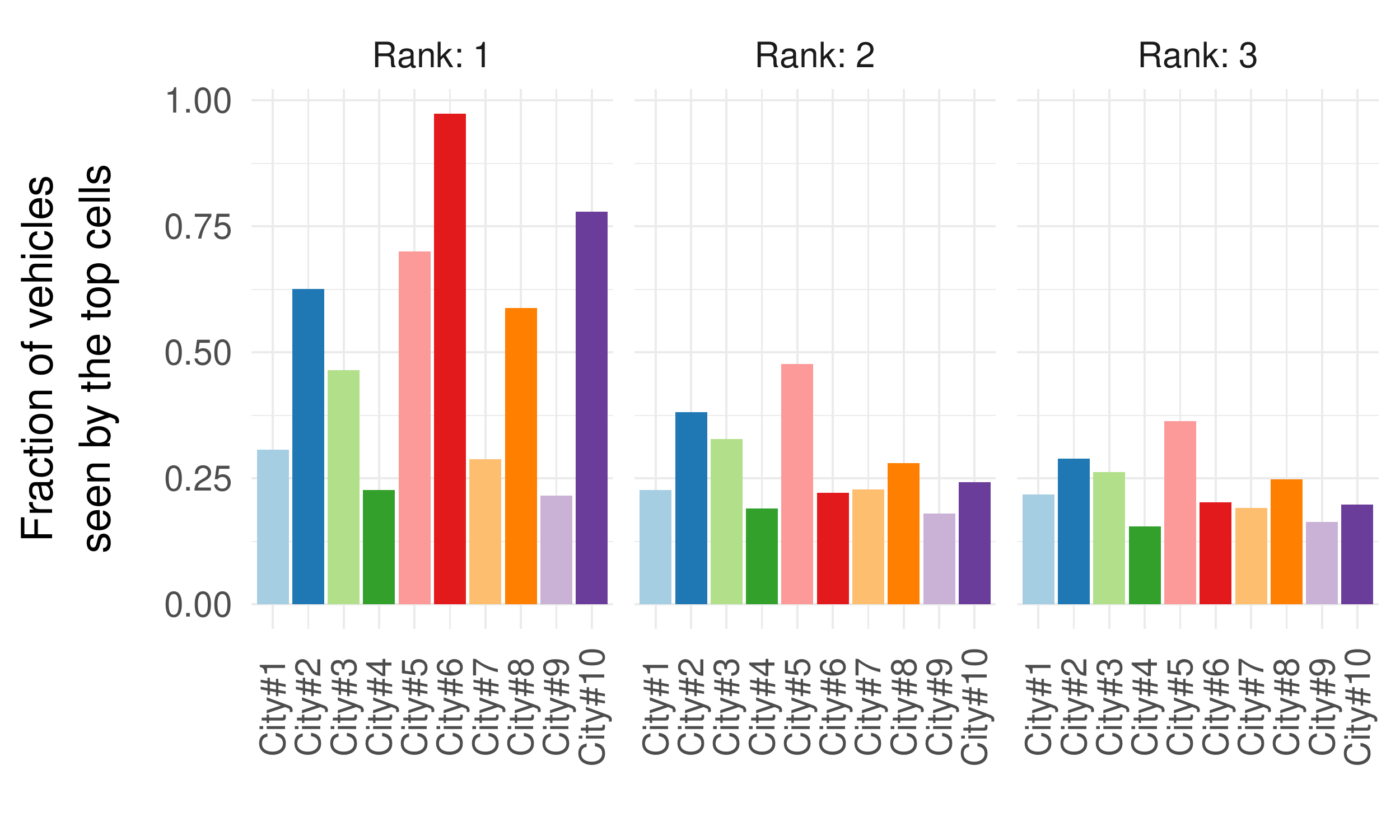}\label{fig:service_areas_w30}}\\
\subfloat[Tolerance window: 15 days]{\includegraphics[scale=0.4]{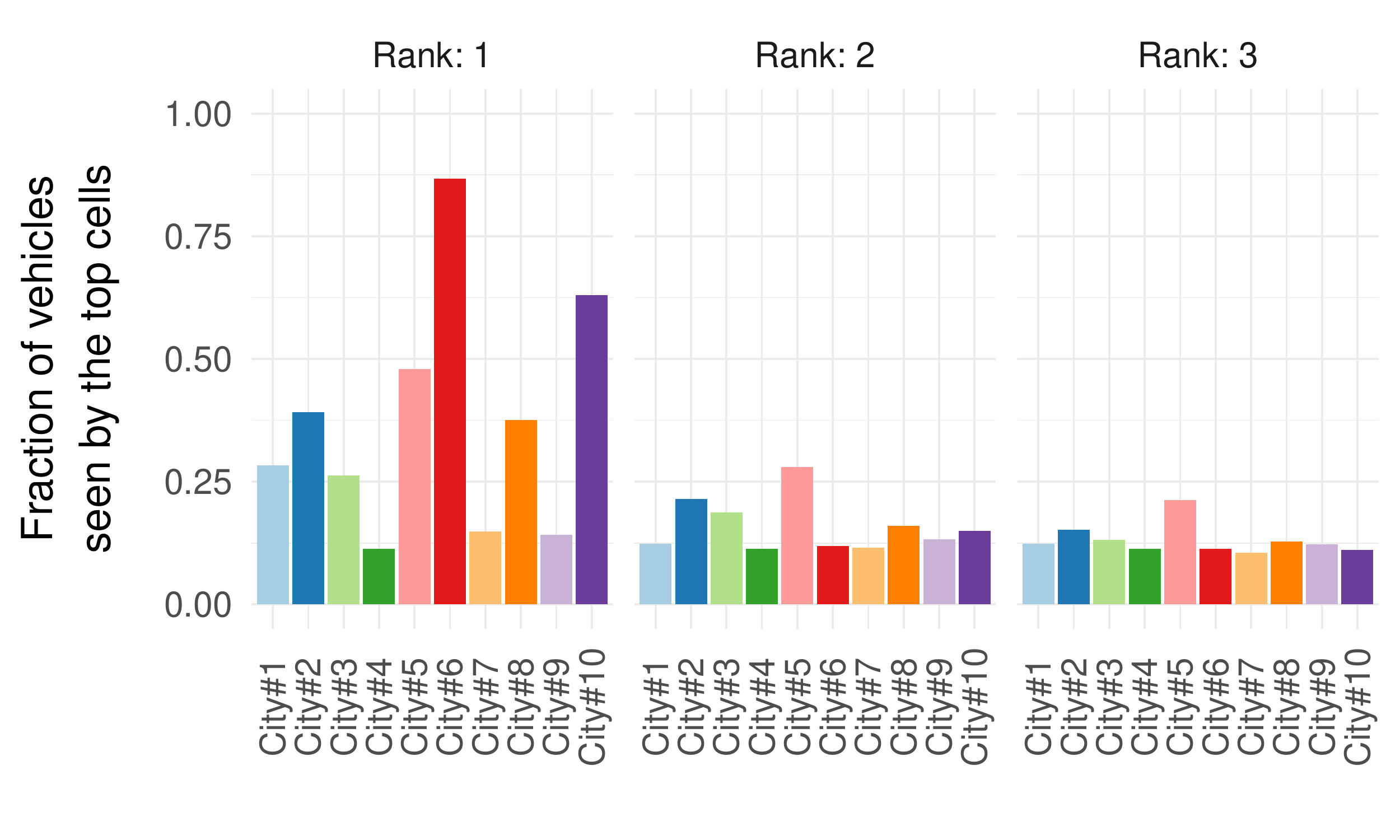}\label{fig:service_areas_w15}}
\end{center}
\vspace{-20pt}
\end{figure}

We can use our dataset to understand if these potential service areas exist or not in the cities covered by the car sharing service under study. To this aim, we define a reference window $W$, corresponding to the accepted tolerance for taking out a vehicle for maintenance. Then, for each cell, we count the number of distinct vehicles seen by the cells during $W$. Figures~\ref{fig:service_areas_w30} and~\ref{fig:service_areas_w15} show the results for the top three cells in each cities, i.e., the three cells that see the highest number distinct vehicles during two different time windows ($W=30$ and $W=15$ days, respectively). Assuming that a (somewhat generous) threshold of $50\%$ vehicles would be acceptable for the car sharing operator to justify the opening of a workshop in the area, only 5 cities out of 10 are able to satisfy this requirement when $W=30$ and only 2 out of 10 when $W=15$. In both cases, the service area would be located at the airport.

%
%\subsection{Trips: exploratory analysis}
%\subsection{Energy}
%\subsection{Rental duration}
%\subsection{Revenue from trips}
%\subsection{Trip classifier}

%%=================================================================================
%\section{Discussion: invariants}
%%
%\noindent
%%
%Is the number of trips generated by a city a function of some city properties? E.g., density of PoIs (theory of intervening opportunities? \emph{Noulas et al. 2012 - A Tale of two cities}) Quality of public transport? Diameter of the city?
%
%Graph of PoIs in the city
%
%%=================================================================================
%\section{Discussion: rebalanceability}
%%
%\noindent
%%
%Is it possible to define a measure of rebalanceability of a CS system? Depending on the demand?
%
%%=================================================================================}
%\section{Discussion: predictive models}
%%
%\noindent

%=================================================================================
%\section{Conclusions}

\section*{Acknowledgements}

This work was partially funded by the ESPRIT project. This project has received funding from the \emph{European Union's Horizon 2020 research and innovation programme} under grant agreement No 653395. This work was also partially funded by the REPLICATE project. This project has received funding from the \emph{European Union's Horizon 2020 research and innovation programme} under grant agreement No 691735.

\bibliographystyle{splncs03}
\bibliography{knowme17}
\end{document}